\documentclass{article}
\usepackage{spconf,amsmath,graphicx}
\usepackage{url}

\usepackage{subfigure}
\usepackage{tabularx}
\usepackage{adjustbox}
\usepackage{caption}

\title{TextrolSpeech: A Text Style Control Speech Corpus with Codec Language Text-To-Speech Models}
\name{Shengpeng Ji, Jialong Zuo, Minghui Fang, Ziyue Jiang, Feiyang Chen, Xinyu Duan, Baoxing Huai, Zhou Zhao$^{\dagger} $\thanks{$\dagger$ Corresponding Author: Zhou Zhao, zhaozhou@zju.edu.cn}}

\address{Zhejiang University, Huawei Cloud\\
$\left \{ shengpengji, jialongzuo, minghuifang, zhaozhou \right \}$ @zju.edu.cn }
\begin{document}
%
\maketitle
\begin{abstract}
Recently, there has been a growing interest in the field of controllable Text-to-Speech (TTS). While previous studies have relied on users providing specific style factor values based on acoustic knowledge or selecting reference speeches that meet certain requirements, generating speech solely from natural text prompts has emerged as a new challenge for researchers. This challenge arises due to the scarcity of high-quality speech datasets with natural text style prompt and the absence of advanced text-controllable TTS models. In light of this, 1) we propose TextrolSpeech, which is the first large-scale speech emotion dataset annotated with rich text attributes. The dataset comprises 236,220 pairs of style prompt in natural text descriptions with five style factors and corresponding speech samples. Through iterative experimentation, we introduce a multi-stage prompt programming approach that effectively utilizes the GPT model for generating natural style descriptions in large volumes. 2) Furthermore, to address the need for generating audio with greater style diversity, we propose an efficient architecture called Salle. This architecture treats text controllable TTS as a language model task, utilizing audio codec codes as an intermediate representation to replace the conventional mel-spectrogram. Finally, we successfully demonstrate the ability of the proposed model by showing a comparable performance in the controllable TTS task. Audio samples are available on the demo page \url{https://sall-e.github.io/}.
\end{abstract}
\section{Introduction}
\label{sec:intro}
In recent years, significant advancements have been made in the field of speech synthesis \cite{fastspeech2,valle,megatts}, with an increasing focus on a more challenging task known as controllable Text-to-Speech (TTS). Previous studies in controllable TTS have predominantly employed either reference audio for style transfer \cite{norespeech} or employed different style factors such as speaking rate \cite{speaking}, pitch \cite{fastpitchformant}, and prosody \cite{skerry2018towards} for speech control. However, these approaches necessitate users to provide specific values for style factors which requiring professional acoustic knowledge \cite{speaking,fastpitchformant,skerry2018towards}, or select reference speech \cite{norespeech} that satisfies the desired criteria. These methods are time-consuming and lack user-friendliness. Moreover, the style information derived from the reference speech lacks intuitiveness and interpretability. The effect over these styles is often constrained to the training set of the reference audio, resulting in weak generalization capabilities for unseen styles.

Based on the aforementioned limitations: PromptTTS \cite{prompttts} proposes that it is a preferable choice to achieve style control using a natural language text description. We believe that utilizing natural text descriptions for controlling style in speech is the direction for future development of controllable TTS systems, due to its user-friendliness, generalizability, and interpretability. However, to the best of our knowledge, there is currently no high-quality, large-scale open-source text style prompt speech dataset available for advanced text-controllable TTS models.
In this work, we introduce a novel 330-hour clean text style prompt speech emotion dataset called TextrolSpeech. Each style encompasses 5 style factors and 500 distinct natural language text descriptions. Given the increased demands of controllable TTS systems for speech diversity, we get inspiration from \cite{valle} and propose Salle, which employs discrete tokens \cite{codec} based Residual Vector Quantization (RVQ) instead of conventional mel spectrograms. The tokens in Salle exhibit a hierarchical structure, where tokens from previous quantizers capture acoustic properties such as speaker identity, while consecutive quantizers learn fine-grained acoustic details. Building upon this characteristic, we directly utilize the text style tokens in an autoregressive manner to prompt the generation of the first layer of acoustic tokens. Our contributions can be summarized as follows:

\begin{table*}[t]
    \centering
    \caption{Comparison between Text style prompt TTS datasets.}
    \begin{tabular}{c|c|ccccc}
        \hline
        Dataset & Open source & Hours & Text items & Prompt diversity  & Speaker & Emotion\\
        \hline
        StylePrompt \cite{promptstyle} & no & 12 & - & - & 8 & -\\
         NLSpeech \cite{instructtts} & no & 44 & 32000 & - & 7 & \textbf{yes}\\
       PromptSpeech \cite{prompttts} & \textbf{yes (part)} & - & 27893 & 5 & - &no\\
        \textbf{TextrolSpeech} & \textbf{yes} & \textbf{330} & \textbf{236220} & \textbf{500} & \textbf{1324} & \textbf{yes}\\
        \hline
    \end{tabular}
    \label{table1}
\end{table*}

\begin{itemize}
    \item We have released TextrolSpeech, an open-source speech emotion dataset that is large-scale, multi-speaker, and enriched with diverse and natural text descriptions. This dataset aims to drive the development of text controllable TTS systems.
    \item We provide a detailed account of the creation process of TextrolSpeech, through our experiments, we have devised an efficient prompt programming methodology that significantly expands the number of text descriptions for each style group.
    \item We propose Salle, a multi-stage discrete style token-guided control framework for TTS language models, which exhibits powerful in-context capabilities.
\end{itemize}

\section{TextrolSpeech Dataset}

\subsection{Overview of the TextrolSpeech}
\label{data1}
Given the absence of large-scale and high-quality TTS datasets with text style prompts, we have created and made available a dataset called TextrolSpeech, which consists of speech samples paired with corresponding style prompts. As shown in Table \ref{table1}, to the best of our knowledge, there have been some previous attempts to construct similar datasets. However, these datasets are often either proprietary \cite{promptstyle,instructtts} or only a small portion of the data is publicly released \cite{prompttts} . In previous works, dataset sizes were also limited to a few tens of hours or several thousand corresponding text prompts, greatly restricting the performance of the models. In contrast, TextrolSpeech provides an open-source dataset consisting of 330 hours of speech data and 236,220 naturally occurring text style descriptions. The speech samples and their corresponding prompts describe the same style and content, and the audio waveforms are resampled to a codec format with a sampling rate of 24kHz. Each speech sample in TextrolSpeech includes five different style factors: gender, pitch, speaking speed, volume, and emotion. The emotion factor comprises eight categories \footnote{angry/contempt/disgusted/fear/happy/sad/surprised/neutral}, while the gender factor has two categories (male/female). The remaining factors, pitch, speaking speed, and volume, consist of three categories (high/low/normal). We randomly set aside 200 samples as the validation set, another 200 samples as the test set, and the remaining data for training purposes.

\begin{figure}
  \centering
  \subfigure[example1]{
    \includegraphics[width=0.22\textwidth]{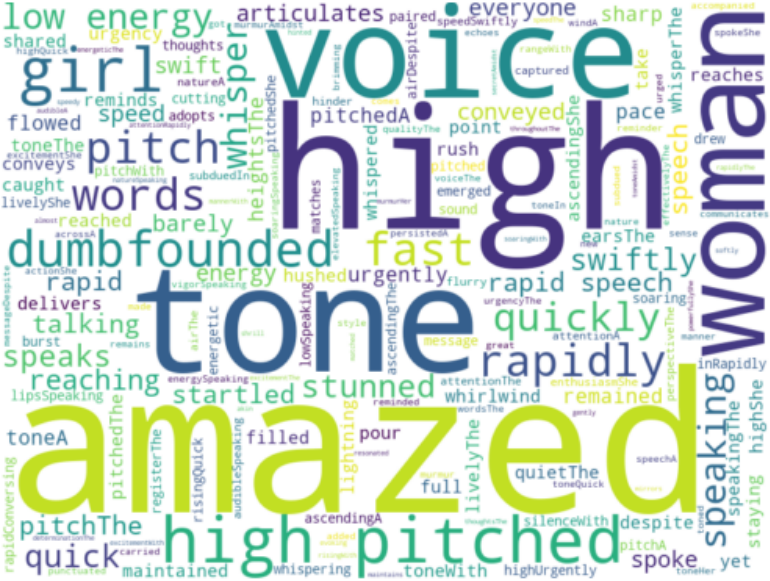}
    \label{fig:subfig1}
  }
  \hfill
  \subfigure[example2]{
    \includegraphics[width=0.22\textwidth]{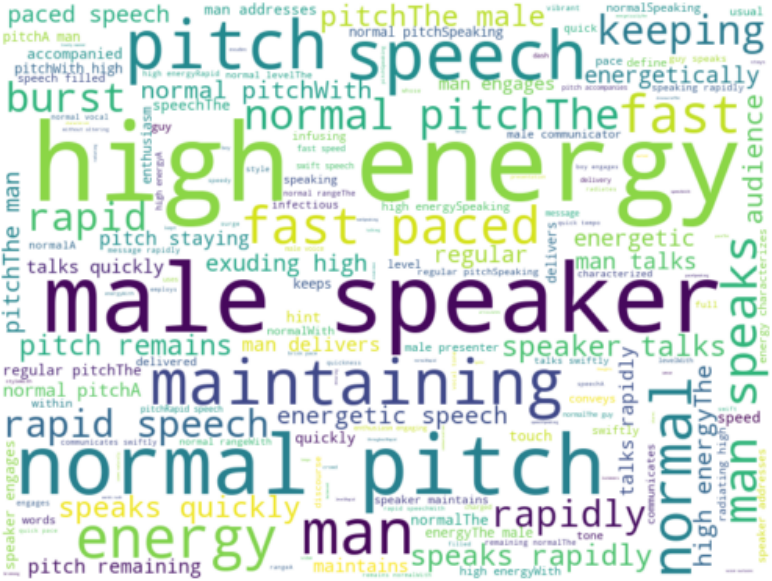}
    \label{fig:subfig2}
  }
  \caption{Two sets of word cloud exemplars}
  \label{fig:subfigures}
\end{figure}

\subsection{Data Collection and Text preprocessing}
\label{data2}
Based on the previous methodology used in PromptTTS \cite{prompttts} for dataset creation, we integrated the clean portion of the LibriTTS dataset \cite{libritts} and the VCTK dataset \cite{vctk} to form a part of the audio corpus in TextrolSpeech. Considering the limited emotional content in these two datasets, we collected and curated a separate set of emotion datasets, namely ESD \cite{esd}, TESS \cite{tess}, MEAD \cite{mead}, SAVEE \footnote{\url{http://kahlan.eps.surrey.ac.uk/savee}} and MESS \cite{mess}. In total, we obtained 42,909 instances of emotional speech data. For those instances lacking emotional content, we labeled their style factors as "neutral." To extract acoustic features, we employed librosa to analyze energy levels and a world vocoder \cite{world} to extract pitch information. Additionally, we aligned the text and speech data to obtain duration information, which was subsequently averaged. Finally, the energy levels, pitch, and speech rate were categorized into three levels (high/low/normal) based on the overall distribution of the data.



\begin{figure*}[t]
\centering
\includegraphics[width=\textwidth]{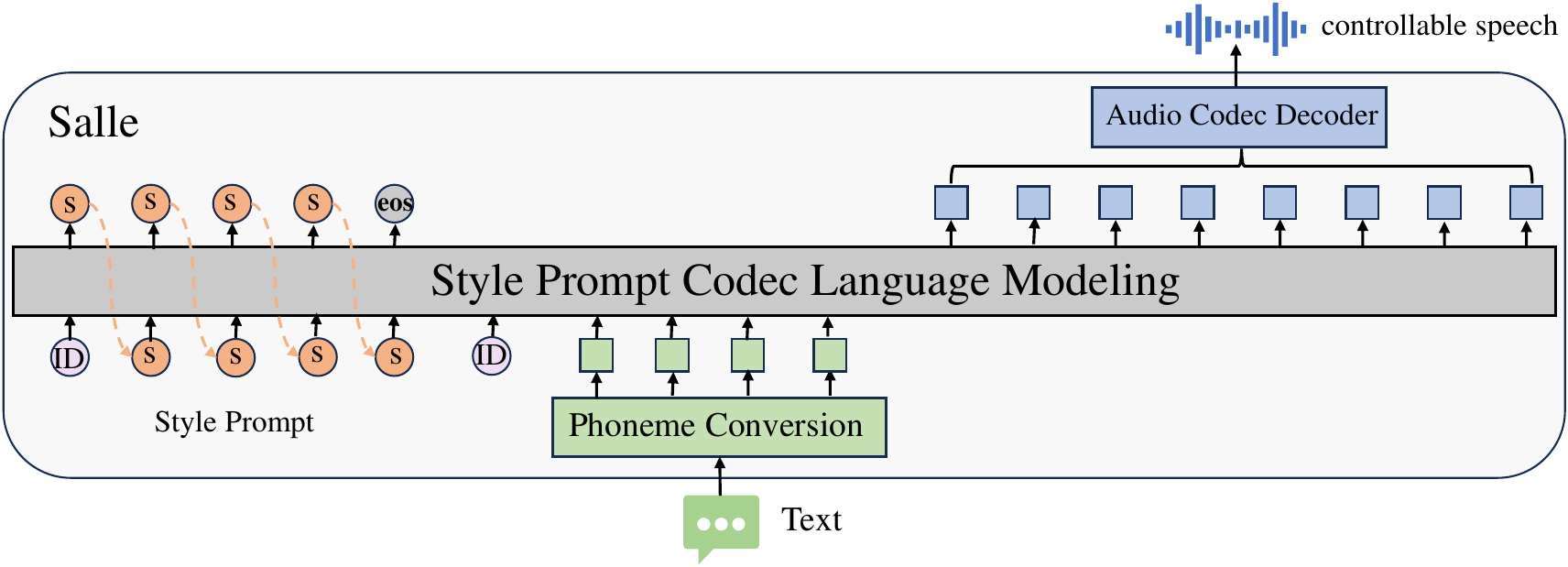} 
\caption{The overall architecture of Salle}
\label{arc1}
\end{figure*}

\subsection{Prompt Programming}
\label{data3}
It is worth noting that we have greatly expanded the diversity of text style prompts through prompt programming. As illustrated in Figure \ref{table1}, we found that previous work had only 5 types of descriptions for each style, which makes it difficult to cover all situations in real-world scenarios and increases the risk of the model learning biases. Through our continuous experimentation, we have designed an efficient multi-stage prompt programming approach. After manual selection, we obtained 500 naturally described prompts for each style group. We utilize GPT-3.5-TURBO to generate style prompts. it is beneficial to consider prompt programming from the perspective of constraint behavior \cite{ppfllm}. We have categorized the prompts provided to GPT into four stages.
\\
\textbf{Base} 
To make it clear to the GPT what needs to be accomplished, we requested in the header of the prompt that the GPT generate text describing the speaking style using style factors that we preprocessed previously. E.g. \textit{Generate one sentence that describe different, natural and brief speaking style based on four keywords: "male", "high pitch", "fast speaking speed", "normal energy".}
\\
\textbf{Increasing diversity}
 We hope that the style prompts generated by GPT will contain the content of the keywords while ensuring diversity, so we allow GPT to use the keywords themselves or their synonyms, or even to generate style prompts based on their own experience. E.g. \textit{you can use them themselves or replace them with their synonyms, and even use some adjectives to describe them. For example, you can use "tone", "key" or "volume" to describe "pitch".}
\\
\textbf{Reducing irrelevant descriptions}
GPT trained on web-scale corpora are very powerful in natural language understanding and reasoning , so it is necessary to limit the GPT to prevent the generation of too much content unrelated to keywords from affecting the quality of style prompts. E.g. \textit{Please remember do not include scene descriptions such as "churches" in the sentences.}
\\
\textbf{Few-shot templates}
The GPT can be conditioned with a few-shot prompts to fully utilize the knowledge it already has\cite{clipcap}. Based on this experience, we have used hand-crafted templates as part of the prompts to enhance the quality of the final generated style prompts. E.g. \textit{I provide you with one correct template based on four keywords that you can refer to and generate diverse sentences. The template is "The rapid, loud and high-keyed voice belongs to the girl".}  Figure \ref{fig:subfigures} shows two sample word clouds for stylistic descriptions. On our demo website, we showcase a collection of 500 distinct descriptions for a group of styles.

\section{Salle}
Based on \cite{codec}, Salle (denoted as $\theta $) leverages style prompt autoregressive codec LM and a non-autoregressive text-to-speech codec LM to generate acoustic tokens at different granularities. Given the dataset $D=\left \{ s_{i}, x_{i}, y_{i} \right \}$, where $(x,y)$ represents a paired set of text and speech, $S=\left \{s_{1},s_{2},s_{3},\cdots,s_{L}  \right \} $ denotes the corresponding text style description, we first get a discrete representation $A^{T*n}$ of each speech y:

\begin{equation}
    A^{T*n}=encodec(y) 
\end{equation}
where $A$ represents the two-dimensional acoustic code matrix, $T$ is the downsampled utterance length and $n$ represents the quantity codes for frame $t$. In all of our experiments, the hyperparameter $n$ is set to 8. 

Taking into consideration the hierarchical structure of the acoustic code, where the first layer codeblock often contains primary speaker information, while the consecutive $n-1$ layers of quantizers learn fine acoustic details. We divide the model into two parts: for the generation of the first layer codeblock $A_{(1:T,1)} $ of speech, we design an autoregressive structure. As shown in Figure 1, the overall codec language $\theta _{SAR} $ is a decoder-only architecture that does not require text-speech alignment. It generates acoustic tokens based on the text style prompt $S$ and corresponding text token embeddings $x$, and stops the generation of speech codec by predicting the $<eos>$ token. This process is formulated as:

\begin{equation}
    p(A_{(1:T,1)}|S,x;\theta _{SAR} ) =\prod_{t=0}^{T} p(A_{(t,1)}|A_{(<t,1)},S,x;\theta _{SAR})
\end{equation}
To expedite the overall speed of the Salle, in the second stage, we employ a parallelized language model $\theta _{SNAR}$ to predict the discrete tokens from the second to the last $n$ quantizers, denoted as $A_{(:,2:n)} $ . In this stage, $\theta _{SNAR}$ follows a conventional speech synthesis model, where we exclude the inclusion of text style prompts. Instead, the generation of each subsequent layer of acoustic codeblocks relies solely on the text information $x$ and the acoustic tokens generated by the previous layers, formulated as:

\begin{equation}
    p(A_{(:,2:n)}|x;\theta _{SNAR} ) =\prod_{i=2}^{n} p(A_{(:,i)}|A_{(:,<i)},x;\theta _{SNAR})
\end{equation}
During the training process of $\theta_{SNAR}$, to mitigate the learned biases of the model, we randomly select $A_{(:,i)}$ as the ground truth in each batch. Both the $\theta_{SAR}$ model and the $\theta_{SNAR}$ model share the same transformer architecture, consisting of 6 layers, 16 attention heads, an embedding dimension of 512, a feed-forward layer dimension of 2048, and a dropout rate of 0.1. We calculate the cross entropy loss only at $A_{(1:T,n)}$. Overall, the prediction of $A$ can be modeled as:



\begin{gather}
\begin{split}
    p(A_{(1:T,n)}|S,x;\theta) =  p(A_{(1:T,1)}|S,x;\theta _{SAR}) \cdot\\
    \prod_{i=2}^{n} p(A_{(1:T,i)}|A_{(1:T,<i)},x,\theta _{SNAR})
\end{split}
\end{gather}


\section{Experiments}
\subsection{Experiment setup}
Salle is trained on 4 NVIDIA TESLA V100 32GB GPUs with a batch size of 6k acoustic tokens per GPU for 200k steps. We optimize the models with the AdamW optimizer, warm up the initial learning rate $1\times 10^{-7}$ for the first 32k updates to a peak of $5\times 10^{-4}$, and then linear decay it. We attempt to reproduce the current SOTA model  PromptTTS, and train it on the TextrolSpeech dataset as a baseline for comparison. In order to assess the model's performance, we train two separate classification models for gender and emotion, achieving classification accuracies of 99.2$\%$ and 85$\%$, respectively. For pitch, speech rate, and energy, we obtain labels using a similar method employed in TextrolSpeech. We analyze the Mean Opinion Score (MOS) ratings from 1 to 5 on two dimensions: MOS-Q (Quality, including attributes such as clarity, high-frequency, and prosody) and MOS-S (Speaker similarity with the style prompt).

\subsection{Results}

\subsubsection{Main Results}
Table \ref{table2} presents the accuracy results of PromptTTS and Salle models on five style factors in the TextrolSpeech test set. Based on the findings, we can draw the following conclusions: In gender, pitch, speech, volume, and emotion, Salle surpasses PromptTTS comprehensively, achieving an average accuracy rate of 87.6$\%$. We attribute this improvement to the architecture based on discrete codecs, which enhances the model's diversity. Additionally, we directly guide the acoustic tokens by utilizing text and style prompts in an autoregressive manner. This approach avoids some information loss that occurs when extracting CLS tokens. Furthermore, we observe that the model easily learns gender-related information, while the classification accuracy for emotions tends to be lower. We attribute this to factors such as limited data, interference from neutral emotions, and the inherent difficulty of emotion classification.

\begin{table}[htp]
    \caption{The Acc($\%$) of Prompttts and Salle on style factors.}
    \begin{adjustbox}{max width=0.48\textwidth}
    \begin{tabular}{c|cccccc}
        \hline
        Models & Gender& Pitch &Speed &Volume &Emotion &Mean \\
        \hline
        PromptTTS & 92.5 & 82.5 & 83 & 82 & 71.5 & 82.3 \\
         Salle & 95.5 & 90.5 & 85 & 86 & 81 & 87.6 \\
        \hline
    \end{tabular}
    \end{adjustbox}
    \label{table2}
\end{table}

\begin{table}[htp]
    \centering
    \caption{The results of MOS-Q and MOS-S with 95$\%$ confidence intervals.}
    \begin{tabular}{c|cc}
        \hline
        Models & MOS-Q & MOS-S \\
        \hline
        GT & 4.21 ± 0.09 & 4.25 ± 0.08 \\
        GT-codec & 4.10 ± 0.11 & 4.13 ± 0.09 \\
        PromptTTS & 3.76 ± 0.12 & 3.74 ± 0.09 \\
         Salle & 3.78 ± 0.09 & 3.88 ± 0.10 \\
        \hline
    \end{tabular}
    \label{table3}
\end{table}

\subsubsection{Speech Quality}
To assess the perceptual quality and similarity, we conducted MOS-Q and MOS-S analysis on four different conditions: 1) GT (Ground Truth), representing the original unaltered audio; 2) GT-codec, where the original audio was encoded and subsequently decoded for reconstruction; 3) PromptTTS, the current state-of-the-art model; and 4) Salle, we proposed model. By referring to Table \ref{table3}, we can observe the following findings:
In subjective MOS experiments, it is evident that Salle outperforms PromptTTS in terms of audio quality and audio similarity. Particularly, the improvement in audio similarity is by 0.14. However, we note that the overall sound quality is not exceptionally high. We attribute this to factors such as limited data and the inherent difficulty of the text style-controlled speech synthesis task itself. We have extensively demonstrated the diversity of results achieved by Salle in our demo.

\section{Conclusion}
 In this paper, we release TextrolSpeech, a large-scale and high-quality text style prompt speech emotion corpus in the field of controllable TTS for the first time. Additionally, we propose a straightforward codec language model named Salle, which is built upon TextrolSpeech. We believe that there is still significant space for improvement in text style control, particularly in enhancing both audio quality and robustness in style control, especially in the domain of emotional effects. We anticipate that TextrolSpeech and Salle can serve as future baselines in this regard.

\vfill\pagebreak



\bibliographystyle{IEEEbib}
\bibliography{refs}

\begin{thebibliography}{10}

\bibitem{fastspeech2}
Yi~Ren, Chenxu Hu, Xu~Tan, Tao Qin, Sheng Zhao, Zhou Zhao, and Tie-Yan Liu,
\newblock ``Fastspeech 2: Fast and high-quality end-to-end text to speech,''
\newblock {\em arXiv preprint arXiv:2006.04558}, 2020.

\bibitem{valle}
Chengyi Wang, Sanyuan Chen, Yu~Wu, Ziqiang Zhang, Long Zhou, Shujie Liu, Zhuo
  Chen, Yanqing Liu, Huaming Wang, Jinyu Li, et~al.,
\newblock ``Neural codec language models are zero-shot text to speech
  synthesizers,''
\newblock {\em arXiv preprint arXiv:2301.02111}, 2023.

\bibitem{megatts}
Ziyue Jiang, Yi~Ren, Zhenhui Ye, Jinglin Liu, Chen Zhang, Qian Yang, Shengpeng
  Ji, Rongjie Huang, Chunfeng Wang, Xiang Yin, et~al.,
\newblock ``Mega-tts: Zero-shot text-to-speech at scale with intrinsic
  inductive bias,''
\newblock {\em arXiv preprint arXiv:2306.03509}, 2023.

\bibitem{norespeech}
Dongchao Yang, Songxiang Liu, Jianwei Yu, Helin Wang, Chao Weng, and Yuexian
  Zou,
\newblock ``Norespeech: Knowledge distillation based conditional diffusion
  model for noise-robust expressive tts,''
\newblock {\em arXiv preprint arXiv:2211.02448}, 2022.

\bibitem{speaking}
Jae-Sung Bae, Hanbin Bae, Young-Sun Joo, Junmo Lee, Gyeong-Hoon Lee, and
  Hoon-Young Cho,
\newblock ``Speaking speed control of end-to-end speech synthesis using
  sentence-level conditioning,''
\newblock {\em arXiv preprint arXiv:2007.15281}, 2020.

\bibitem{fastpitchformant}
Taejun Bak, Jae-Sung Bae, Hanbin Bae, Young-Ik Kim, and Hoon-Young Cho,
\newblock ``Fastpitchformant: Source-filter based decomposed modeling for
  speech synthesis,''
\newblock {\em arXiv preprint arXiv:2106.15123}, 2021.

\bibitem{skerry2018towards}
RJ~Skerry-Ryan, Eric Battenberg, Ying Xiao, Yuxuan Wang, Daisy Stanton, Joel
  Shor, Ron Weiss, Rob Clark, and Rif~A Saurous,
\newblock ``Towards end-to-end prosody transfer for expressive speech synthesis
  with tacotron,''
\newblock in {\em international conference on machine learning}. PMLR, 2018,
  pp. 4693--4702.

\bibitem{prompttts}
Zhifang Guo, Yichong Leng, Yihan Wu, Sheng Zhao, and Xu~Tan,
\newblock ``Prompttts: Controllable text-to-speech with text descriptions,''
\newblock in {\em ICASSP 2023-2023 IEEE International Conference on Acoustics,
  Speech and Signal Processing (ICASSP)}. IEEE, 2023, pp. 1--5.

\bibitem{codec}
Alexandre D{\'e}fossez, Jade Copet, Gabriel Synnaeve, and Yossi Adi,
\newblock ``High fidelity neural audio compression,''
\newblock {\em arXiv preprint arXiv:2210.13438}, 2022.

\bibitem{promptstyle}
Guanghou Liu, Yongmao Zhang, Yi~Lei, Yunlin Chen, Rui Wang, Zhifei Li, and Lei
  Xie,
\newblock ``Promptstyle: Controllable style transfer for text-to-speech with
  natural language descriptions,''
\newblock {\em arXiv preprint arXiv:2305.19522}, 2023.

\bibitem{instructtts}
Dongchao Yang, Songxiang Liu, Rongjie Huang, Guangzhi Lei, Chao Weng, Helen
  Meng, and Dong Yu,
\newblock ``Instructtts: Modelling expressive tts in discrete latent space with
  natural language style prompt,''
\newblock {\em arXiv preprint arXiv:2301.13662}, 2023.

\bibitem{libritts}
Heiga Zen, Viet Dang, Rob Clark, Yu~Zhang, Ron~J Weiss, Ye~Jia, Zhifeng Chen,
  and Yonghui Wu,
\newblock ``Libritts: A corpus derived from librispeech for text-to-speech,''
\newblock {\em Interspeech 2019}, 2019.

\bibitem{vctk}
Christophe Veaux, Junichi Yamagishi, Kirsten MacDonald, et~al.,
\newblock ``Cstr vctk corpus: English multi-speaker corpus for cstr voice
  cloning toolkit,''
\newblock {\em University of Edinburgh. The Centre for Speech Technology
  Research (CSTR)}, vol. 6, pp. 15, 2017.

\bibitem{esd}
Kun Zhou, Berrak Sisman, Rui Liu, and Haizhou Li,
\newblock ``Seen and unseen emotional style transfer for voice conversion with
  a new emotional speech dataset,''
\newblock in {\em ICASSP 2021-2021 IEEE International Conference on Acoustics,
  Speech and Signal Processing (ICASSP)}. IEEE, 2021, pp. 920--924.

\bibitem{tess}
Kate Dupuis and M~Kathleen Pichora-Fuller,
\newblock ``Toronto emotional speech set (tess)-younger talker\_happy,''
\newblock 2010.

\bibitem{mead}
Kaisiyuan Wang, Qianyi Wu, Linsen Song, Zhuoqian Yang, Wayne Wu, Chen Qian, Ran
  He, Yu~Qiao, and Chen~Change Loy,
\newblock ``Mead: A large-scale audio-visual dataset for emotional talking-face
  generation,''
\newblock in {\em European Conference on Computer Vision}. Springer, 2020, pp.
  700--717.

\bibitem{mess}
Shae~D Morgan,
\newblock ``Categorical and dimensional ratings of emotional speech: Behavioral
  findings from the morgan emotional speech set,''
\newblock {\em Journal of Speech, Language, and Hearing Research}, vol. 62, no.
  11, pp. 4015--4029, 2019.

\bibitem{world}
Masanori Morise, Fumiya Yokomori, and Kenji Ozawa,
\newblock ``World: a vocoder-based high-quality speech synthesis system for
  real-time applications,''
\newblock {\em IEICE TRANSACTIONS on Information and Systems}, vol. 99, no. 7,
  pp. 1877--1884, 2016.

\bibitem{ppfllm}
Laria Reynolds and Kyle McDonell,
\newblock ``Prompt programming for large language models: Beyond the few-shot
  paradigm,''
\newblock in {\em Extended Abstracts of the 2021 CHI Conference on Human
  Factors in Computing Systems}, 2021, pp. 1--7.

\bibitem{clipcap}
Ron Mokady, Amir Hertz, and Amit~H Bermano,
\newblock ``Clipcap: Clip prefix for image captioning,''
\newblock {\em arXiv preprint arXiv:2111.09734}, 2021.

\end{thebibliography}

\end{document}